\documentclass{elsart}
\usepackage{graphicx}
\usepackage{amssymb}
\input{epsf}
\parskip=\medskipamount            
  \def\CO{{\cal O}}

\def\be{\beta}

\def\Bbb{\mathbb}

\font\af=msbm10
\def\R{{\Bbb R}}      
\def\uno{\relax{\rm 1\kern-.25 em l}}

\def\IK{\relax{\rm l\kern-.18 em K}}
\def\IL{\relax{\rm I\kern-.18 em L}}
\def\IR{\hbox{\af R}}

\def\Id{\mathop{\rm Id}\nolimits}

\def\pd#1#2{\frac{\partial #1}{\partial#2}}

\def\wt{\widetilde}
\def\frac#1#2{{#1\over #2}}

\def\ptos{\leaders\hbox to 2mm{\hfil{.}\hfil}\hfill}
\def\\{\hfill\break}

\def\<#1>{\langle#1\rangle}
\font\tenfrak=eufm10  \font\sevenfrak=eufm7  \font\fivefrak=eufm5
\newfam\frakfam
\textfont\frakfam=\tenfrak\scriptfont\frakfam=\sevenfrak
\scriptscriptfont\frakfam=\fivefrak

\font\tengoth=eufm10 scaled\magstep1 \font\sevengoth=eufm7 \font\fivegoth=eufm5
\newfam\gothfam
\textfont\gothfam=\tengoth\scriptfont\gothfam=\sevengoth
  \scriptscriptfont\gothfam=\fivegoth

\def\today{\ifcase\month\or
   January\or February\or March\or April\or May\or June\or
   July\or August\or September\or October\or November\or December\fi
   \space\number\day, \number\year}
\def\be{\begin{equation}}
\def\ee{\end{equation}}
\def\beno{\begin{eqnarray*}}
\def\eeno{\end{eqnarray*}}

\begin{document}
\begin{frontmatter}
\title{ \bf  Isoperiodic classical systems and their quantum counterparts}
\author[uni]{M. Asorey}, \author[uni]{J. F. Cari\~nena},\author[due,trei]{G.
Marmo}, \author[four]{A. Perelomov}
 \address[uni]{   {\it Departamento de F\'{\i}sica Te\'orica, 
Facultad de Ciencias}
             {\it Universidad de Zaragoza, 50009 Zaragoza, Spain}}
$^{\ddagger}$
  \address[due]{             {\it Dipartimento  di Scienze Fisiche, Universit\'a
Federico II di Napoli}}
 \address[trei]{  
and {\it INFN, Sezione di Napoli},
             {\it Complesso Univ. di Monte Sant'Angelo, Via Cintia, 80125
Napoli, Italy}}
\address[four]{  {\it Institute for Theoretical and Experimental 
Physics, 117259 Moscow, Russia}.
}

\begin{abstract}


 One-dimensional isoperiodic classical  
systems have been  first analyzed  by 
Abel. Abel's  characterization can be extended for 
singular potentials and potentials which are not defined on
the whole real line. The standard shear equivalence of
isoperiodic potentials can also be extended by using reflection and inversion 
transformations. We provide a full characterization of isoperiodic rational potentials showing that they
are connected by translations,  reflections or Joukowski transformations.
Upon quantization  many of these isoperiodic systems fail to exhibit identical quantum energy spectra.
This  anomaly occurs at order $\CO(\hbar^2)$ because   semiclassical corrections of
energy levels of order  $\CO(\hbar)$ are identical for all isoperiodic systems. 
We analyze families of systems where this quantum anomaly occurs and some 
special systems where the spectral identity is preserved by
quantization.
Conversely, we point out the existence of  isospectral quantum systems which 
do not correspond to isoperiodic classical systems.
\bigskip

\end{abstract}

\end{frontmatter}


{\vfill}
{\small
}
\newpage

\section{Introduction}

The connection between classical and quantum physics has always been
tantalizing and elusive. The establishment of 
quantization rules for classical system has been
the algorithmic method which dominated  the construction of quantum systems.
This pathway has been plagued with surprises: existence of quantum 
anomalies, operator ordering problems, quantum divergences, 
spontaneous symmetry breaking, renormalization of couplings and observables, etc.
The way back to classical mechanics from quantum dynamics has revealed 
also  problematic due to the failure of semiclassical expansion and the
existence of quantum states without a natural classical analogue.
One of the most explicit realizations of the genuine differences between
classical and quantum systems is provided by the analysis of boundary
conditions in systems evolving in constrained spaces \cite{aim,aim2}.
However, this mismatch has been very useful to  introduce new  quantum inspired
classical structures: quantum groups, non-commutative geometry, etc.
 
In this note we explore the analogies and differences between the equivalences
of classical and quantum systems from a spectral point of view.
There is a natural equivalence relation between classical mechanical systems
based on the analysis of the periods of closed orbits and its dependence on the
orbit energy. Two bounded mechanical systems might be considered equivalent 
if they employ the same time periods for closed orbits with the same  energy,
and then they are said to be isoperiodic. 
This equivalence relation was introduced by Abel in 1826 \cite{Ab26}.
 It can be shown that the equivalence
classes of equivalent  potentials include potentials
 related by  shear transformations but this does not exhaust all possibilities
 as we will show below. This fact is related with 
 the concept of Steiner symmetrization which was used in \cite{Steiner} to 
establish that all potentials with only one minimum having the same
  Steiner symmetrized potential have the same dependence of periods on the
energy $T(E)$.

Another open problem is the characterization of
all potentials which give rise to isochronous motions, i.e. the period does not
depend on the energy (see e.g. \cite{{BMK03},{calog}}, and references therein). 
The origin of the problem
is even older, it goes back to Huygens in 1673 \cite{huygens}. In the one-dimensional case
with rational potentials it can be shown that the  only symmetric isochronous potentials
with a constant period $T=2\pi/ \omega$ correspond either to the harmonic
oscillator $U(x)=\frac 12 \omega^2 x^2$ or 
to the isotonic potential  of the form
 $U(x)=\frac 18 \omega^2 x^2+\alpha x^{-2}$,  up  to a translation \cite{vesel}. 

 There is a similar equivalence relation for quantum systems. Two
quantum systems with  bounded classical analogs  are said to be spectrally equivalent if their  energy levels 
 are identical. It is well known  that the isoperiodicity equivalence is the classical version
of the quantum isospectrality condition (see e.g. \cite{EKK97} and \cite{JD05} 
 for a recent
discussion). In the same way it is obvious that the quantum counterpart of 
isochronicity is the harmonic spectrum (for regular potentials). 
However, there is not a theorem characterizing the
 potentials with equally spaced energy levels in an analogous way
as for the  isochronous systems. In particular, we shall show 
the existence of   many 
isochronous classical systems which do not have equally spaced quantum energy
 spectra.

More generally, the classical equivalence associated to isoperiodicity is not
always preserved by the quantization process, i.e. given two isoperiodic
classical systems the corresponding quantum 
systems might not be isospectral. The exploration of the anomalies of this 
correspondence is one of the goals
of this paper. In particular, we will exhibit
  many isochronous classical systems which do not have equally spaced 
quantum energy spectra. Conversely, we will also show that there are spectrally equivalent quantum systems 
which are not classically isoperiodic.

In the path integral approach to quantum mechanics the anomaly can be understood by the
simple fact that the contribution of paths which do not correspond to classical
solutions of motion equations give different contributions for some isoperiodic
potentials. However,  the equivalence between isoperiodic classical
systems is not broken in the semiclassical approximation  $\CO(\hbar)$.
Thus, the quantum anomalies, when  they exist, can only appear in higher
order corrections $\CO(\hbar^2)$.

Isoperiodic  deformations of an  potential
 are of two types: shear transformations  and time-space scale transformations.
The difference between both deformations is that 
in the first case the  energy of the orbits is  preserved whereas in 
the second case the
energy levels are scaled. The quantum anomaly in the first case can be 
interpreted
as an obstruction to the shear transformation which requires an additional
amount of energy to be performed unlike for  the classical systems. On the
 contrary,
 the scale transformation always involves energy transfer in both cases.
One of the main results of the paper is the proof that the full characterization of isoperiodic 
rational potentials can be achieved in terms of translations, reflections and Joukowski transformations.

On the other hand, there are quantum mechanical systems whose potentials are 
related by a Darboux transformation that implies
that they have (almost) identical energy spectra. Some of those systems turn out to be
classically isoperiodic but some others do not. These facts illuminate the 
relations existing between quantum isospectrality and classical
isoperiodicity, two similar but not identical dynamical concepts.
 The analysis of these equivalences at the classical and quantum levels
provides  a very illuminating picture of the quantum/classical transition.

The paper is organized as follows:  In Section 2 we analyze the notion
of isoperiodicity and provide a characterization of polynomial isochronous
potentials. The generalization of Abel's theory for singular potentials is
approached in Section 3, where we prove the main results of the paper
concerning the characterization of isoperiodic rational potentials 
which are illustrated  with
some illuminating examples. In  Section 4 we analyze the role of scale invariance 
in the analysis of isoperiodicity.  The quantum analogue of isoperiodicity is isospectrality.
The appearance of anomalies in the quantization of isoperiodic potentials prevents  the 
isospectrality of  some isoperiodic potentials, although their  first order semiclassical corrections
are identical. This is shown in Section 5 while in Section 6
we analyze the opposite case, where we analyze some isospectral potentials related by Darboux transforms
which are not classically isoperiodic.

\section{Isoperiodic Potentials}

The complete identification of all potentials of one-dimensional mechanical systems giving
rise to the same dependence $T(E)$ 
of the period $T$ of recurrent trajectories with an energy $E$ was provided by
Abel \cite{Ab26} (see also \cite{LL76}). This classification can also be
  obtained by identifying the
deformations of the potential that do not change the $T(E)$ dependence.
For one dimensional problems this period/energy dependence is given by 
\be
T(E)=\sqrt{ 2\,m}\int_{x_m(E)}^{x_M(E)}\frac{dx}{\sqrt{E-U(x)}}\ .
\label{period}
\ee
where 
$m$ is the mass of the particle and $x_m(E)$ and $x_M(E)$ denote the  two turning points  which are the roots of
the equation $U(x)=E$.

Here $U(x)$ will be  assumed to be a convex potential of the form displayed in Figure
1. Having in mind the invariance under translations, we can assume in the simplest
case, the asymptotic behavior
$\lim_{x\to\pm \infty}U(x)=\infty$ and that
$U$ has  two branches,

\begin{eqnarray*}
U(x)=\left\{\begin{array}{cc} U_1(x)&{\rm if\ } x<0\cr
U_2(x)&{\rm if\ } x>0\end{array}\right.
\end{eqnarray*}

where  $U_1(x)$ and $U_2(x)$ are two monotone decreasing/increasing
functions, i.e. such that $x\,U'(x)>0$. The inverse maps of $U_1(x)$ and $U_2(x)$ will be denoted $x_1(U)$ and
$x_2(U)$, respectively. Note that their values for $U=E$ are those of the
turning points.

 For more general non-convex potentials like those of
Figure 2 the period does not depend only on the energy $E$ but also depends on the periodic branch
specified by $x_m(E)$ and $x_M(E)$, i.e. $T(E,x_m,x_M)$. Note that such type
potentials cannot be isochronous \cite{vesel}.

We shall restrict
ourselves in this section to the convex case and we shall postpone  the
discussion of other interesting  cases, for instance those
in which the potential  presents poles, for next sections. The existence of nodes splits the one-dimensional space
into isolated domains bounded by the poles where the dynamics of the system  is confined.

\begin{figure}
\hskip4cm\includegraphics[width=.6\textwidth]{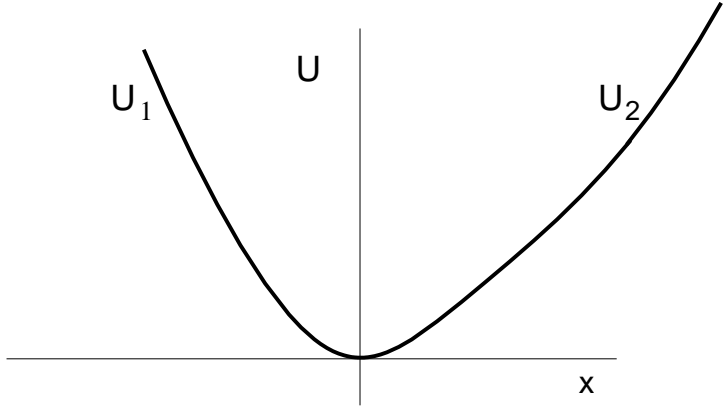}
\caption{Generic convex potential}
\end{figure}

It was shown by Abel \cite{Ab26} that the relation between energy and period 
given by (\ref{period})
does not uniquely determine the potential $U$, but only the difference
 $x_2(U)-x_1(U)$, which is given by:
\be
x_2(U)-x_1(U)=\frac 1{\pi\, \sqrt{2\,m}}
\int_0^U\frac {T(E)}{\sqrt{U-E}}\, dE\ .
\label{invv}
\ee
For a proof using Laplace transformation see e.g. \cite{OO87}.
 This expression shows that the general potential $U(x)$ having a given 
 period/energy
 dependence can be expressed  by means of a particular solution $x^0_i$,
 $i=1,2$,
and 
choosing an arbitrary   function $g:\IR\to \IR$, and in
terms of $g$ the general solution of the Abel equation is
\be
x_2(U)=x^0_2(U)+ g(U),\qquad x_1(U)=x^0_1(U)+ g(U)\,.
\label{gen0}
\ee
Obviously $g$ should not alter the assumed convex character of the potential. 

In particular, if we choose $g(U)=-\frac 12 (x^0_1(U)+x^0_2(U))$ we find a
solution $(x^s_1,x^s_2)$  such that $x^s_1(U)=-x^s_2(U)$, 
\beno
x^s_2(U)=\frac 12 (x_2(U)-x_1(U))\,,\qquad x^s_1(U)=-x^s_2(U)\,,
\eeno
and therefore
corresponding to a potential $U^s$ which is symmetric under reflection with respect
to the origin, i.e. $U^s(-x)=U^s(x)$. Such potential is nothing but the Steiner
symmetrization \cite{Steiner} of the starting potential $U$.
 Using such a
particular solution, 
 the general solution is given by 
\be
x_2(U)=x^s_2(U)+ g(U),\qquad x_1(U)=-x^s_2(U)+ g(U)\,.
\label{gens}
\ee

Under the additional assumption  that the potential $U$ is 
symmetric under  reflection   with respect to the origin there is a unique
potential $U$ with a given period/energy dependence satisfying (\ref{period}),
which will be the one given for $U>0$ by \cite{BFK1,BFK2,{AHC}}:
\be
x^s_2(U)=\frac 1{2\pi\,\sqrt{2\,m}}\int_0^U\frac {T(E)}{\sqrt{U-E}}\, dE\ .
\label{invs}
\ee
\begin{figure}
\hskip4cm\includegraphics[ width=.5\textwidth]{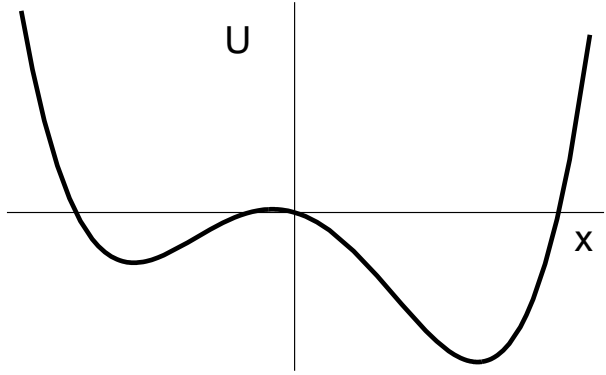}
\caption{Non-convex potential}
\end{figure}
 Now, because of the convex character of $U$ the relation (\ref{invs}) 
can be inverted giving $U^s$ as a function of $x^s_2$ for $x^s_2\geq 0$.

Note that relation (\ref{gen0}) can be inverted giving rise to a relation
\begin{equation}
 U(x)=U^0(x+ f(U(x)))\,,\label{shearrelated}
\end{equation}
and more particularly, when $U^0$ is the symmetric potential $U^s$
in its equivalence  class of 
potentials, we obtain from  (\ref{gens}) the relation 
\beno
U(x)=U^s(x+ g(U(x))\,.
\eeno

There is  a very fundamental identity which characterizes all potentials (related by a
shear transformation) having the $T_U(E)$ dependency  of the period as a
function of the energy corresponding to the potential $U$:
\be
U(x)=U\left(x+W_U(U(x))\right) \qquad 
{\rm where}\qquad
W_U(V)=\frac 1{\pi\, \sqrt{2\,m}}
\int_0^V\frac {T_U(E)}{\sqrt{V-E}}\, dE\ ,
\ee
which is  easily derived from Abel relation (\ref{invv}).

The potentials in the same class of shear equivalence as a given potential $U$ can be characterized as
the fixed points of the following transformation
\be
\widetilde {U}(x)=\bar U\left(x+W_U(\bar U(x))\right).
\label{rg}
\ee
This transformation can be considered as a classical analog of a
renormalization group transformation. When compared with (\ref{shearrelated}),
it shows that it is completely characterized by the
choice of the basic potential $U$ which determines the nature of the fixed point
potentials which are shear equivalent to $U$.

This renormalization group transformation is very useful to further
characterize the isoperiodic potentials within a certain class of potentials.
In particular, is very useful to prove some theorems concerning
isochronous rational potentials.

It is commonly believed that the harmonic oscillator is the only polynomial potential which is
isochronous. This guess can be substantiated in  more rigorous terms \cite{Appell} .

{\bf Theorem 1.} A convex polynomial  potential $U(x)$ is  isochronous
 iff $U(x)=a x^2 +bx +c$

{\it Proof}: If the potential $U(x)$ is  an
isochronous  potential with period $T$ we can 
use (\ref{invv}) when $T(E)$  is  constant and  we obtain
\beno
x_2(U)-x_1(U) =\frac 1{ \pi\, \sqrt{2\,m}}
\int_0^U\frac {T}{\sqrt{U-E}}\, dE\ ={ 2\,T\over \pi \sqrt{2m}}\sqrt{U}\,, 
\eeno
which implies that \cite{vesel}
\be
U(x)=U\left(x+{2T\over\pi \sqrt{2m}}\sqrt{U(x)}\right). 
\label{dd}
\ee
From the analysis of the leading term of (\ref{dd}) we see that 
a   solution $U$ of (\ref{dd})
can be  polynomial only if
  $U$ is the square of a linear polynomial, i.e. $U(x)=(\alpha\,x+\beta)^2$
  which proves the theorem. The condition $U(0)=0$ fixes $\beta=0$
and yields to the standard harmonic oscillator. 

A generalization for the case of rational functions is also
possible and will be considered in next section.

 Let us examine  some 
examples of the isochronous case for which  $T(E)$  takes 
a constant value $T$. 
In that case, using (\ref{invs}) we see that   $x^s(U)=(T/\pi)\sqrt{U/(2m)}$ and the general solutions
  $x_1(U)$ and 
$x_2(U)$ are, respectively \cite{Ab26},
\be
x_1(U)=-{T\over \pi}\sqrt {U\over 2m}+g(U)\,,\qquad x_2(U)={T\over \pi}\sqrt {U\over 2m}+g(U)\ .
\ee

{\bf Case A}. If we choose $g(U)=a$ we find 
\be
x_1(U)=-{T\over \pi}\sqrt {U\over 2m}+a\,,\qquad x_2(U)={T\over \pi}\sqrt {U\over 2m}+a\,,
\ee
from which we obtain the harmonic oscillator  potential centered at $x=a$.
\be
U(x)=
\frac{m\,\omega^2}2\,(x-a)^2\ ,\qquad {\rm with}\quad  \omega=\frac{2\,\pi}T\,.
\ee
{\bf Case B}. If, instead,  the function $g$ is  chosen as $g(U)=\alpha \, 
(T/\pi)\sqrt{U/(2m)}$, then
\be
x_1(U)=(-1+\alpha)\, {T\over \pi}\sqrt {U\over 2m}\,,\quad x_2(U)=(1+\alpha)\,
{T\over \pi}\sqrt {U\over 2m}\ ,
\ee
which for  $|\alpha|\neq 1$, corresponds to the potential of two
half-oscillators \cite{{OO87},{PM00},{GH}}  sometimes called split-harmonic oscillator \cite{SS89,{JD05}}:
\be
U(x)= \cases {\frac 12\,{m\, \omega_1^2\, x^2\  }&{ if  $x\leq 0$ }\cr\cr
\frac 12{m\, \omega_2^2\, x^2\ } & { if $x\geq 0$}}\label{twofreq}
\ee
with different angular frequencies 
\be
\omega_1={2\pi \over ({1-\alpha})T}=\frac{\omega_0}{1-\alpha}\,,\qquad 
\omega_2={2\pi \over ({1+\alpha}) T}=\frac{\omega_0}{1+\alpha}\,,
\ee
glued together at the origin of coordinates \cite{PM00}.
Note that 
\begin{equation}
1/\omega_1+1/\omega_2=2/\omega_0\,,
\label{semiT}
\end{equation}
where $\omega_0=2\pi/T$. The harmonic oscillator with $\omega=\omega_0/2$ is
the Steiner symmetrized of this split-harmonic oscillator \cite{Steiner}.

 Conversely, given
  a potential like in (\ref{twofreq}) we can reduce it to Case B with
the choice \cite{OO87}
\beno
\alpha=\frac{\omega_1-\omega_2}{\omega_1+\omega_2}\qquad {\rm  and}\qquad
\omega_0=\frac{2\,\omega_1\,\omega_2}{(\omega_1+\omega_2)}\,.
\eeno
Note that this potential (\ref{twofreq}) is not analytic at $x=0$ but 
$U''(0+)-U''(0-)=m(\omega^2_2-\omega_1^2)$.

The limit cases $\alpha=\pm 1$ correspond to the half harmonic oscillator and
its reflected one, to be studied later. 
\begin{figure} 

\hskip3cm\includegraphics[width=.7\textwidth]{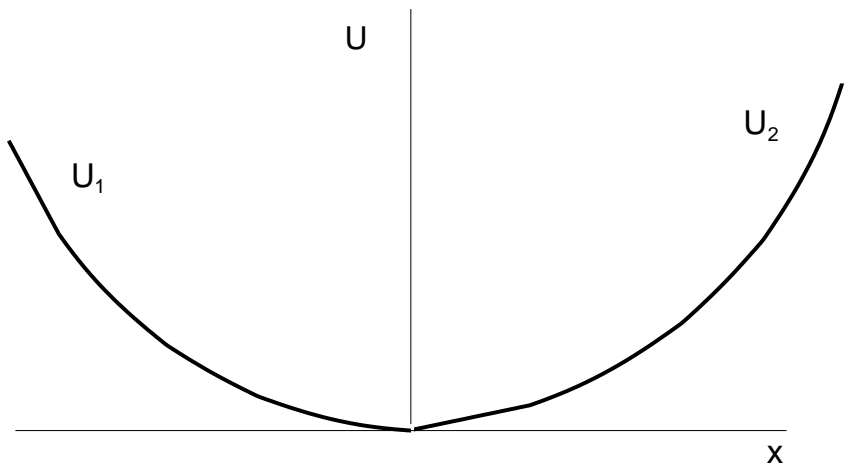}
\caption{ Split--harmonic oscillator with two different frequencies  
$\omega_1,\omega_2$}
\end{figure}

\section{Singular potentials and shear equivalence}
There are two slight generalizations of Abel theorem for
non-convex and singular potentials.
The first  one arises as a consequence of  the Euclidean symmetry 
of  the kinetic term of mechanical systems. In particular, it is invariant under
space translations and reflections. The space translation  symmetry is
the cause for  the ambiguity in isoperiodic systems associated to the 
choice of the shear function $g(U)=a$. Now, the reflection symmetry 
interchanges the order of turning points of closed
trajectories and establishes the mechanical equivalence (isoperiodicity) 
 of  a potential $U$ and its space reflected pair $U^R(x)=U(-x)$, for which 
$x^R_{1}(U)=-x^R_{2}(U)$ and $x^R_{2}(U)=-x^R_{1}(U)$. It is obvious 
that this operation preserves
the relations period/energy for any potential. In the case of convex potentials
the equivalence is included in the Abel family of isoperiodic
solutions. However, for non-convex potentials the reflection transformation 
introduces a new type of solution not included in Abel's family of isoperiodic
potentials. The most general solution for any smooth potential is thus given from a
particular solution $U_\ast$,  its reflected pair $U_\ast^R$ and  their shear equivalents
\be
 U_g(x)=U_\ast(x-g(U_g(x)))\qquad U_g^R(x)=U_\ast(-x+g(U_g(-x)))\,.
\label{invgen}
\ee
The reflection symmetry could be in principle defined with respect to any point
of the real line, but the isoperiodic potentials obtained by this
transformation are included in those of (\ref{invgen}) because the most general
reflection can be expressed as a composition of reflections with respect to the
origin and translations, both considered in (\ref{invgen}).

The second generalization of
  Abel's  solution (\ref{gen0}) concerns the case of  singular potentials 
or potentials which are  not defined on the whole real line. In that case one
has to look for new types of isoperiodic potentials. Let us analyze once again
the isochronous case. In that case we have a generalization of Theorem 1 for
the case of  rational potentials.

{\bf Theorem 2.}  A rational
  potential $U(x)$ which does not reduces to a polynomial  is isochronous
iff \footnote{This theorem was first proved by Chalykh and Veselov \cite{vesel}.
The proof below is a different proof.}
\beno
U(x)=\left(\frac{a x^2 + bx +c }{x+d}\right)^2\,.
\eeno

{\it Proof}: 
 Any rational potential $U(x)$ solution to (\ref{dd})
requires that $U(x)$ has to be  the square
 of the irreducible quotient of two polynomials $P(x)$ and $Q(x)$,
\beno
U(x)=\left(\frac{P(x)}{Q(x)}\right)^2\,.
\eeno
  The stability of the
leading term under the non-linear constraint (\ref{dd}) requires that
the degree of $P$ cannot be higher than one unit more than that of $Q$. As $U(x)$ was
assumed to be rational we can consider the analytic continuation of such a function to
the complex plane.
 If $Q$ is not
constant the potential develops at least one pole in the complex plane which
 is not a zero of $P$  
because $P$ and $Q$ cannot have common zeros.
We will show that $Q(x)$   cannot have two different 
zeros and therefore that in such a case the zero 
 should be real. 
Indeed, if $w$ is a zero of $Q$ let us consider the function $R_w(z)$ given by 
\begin{equation}
 R_w(z)=Q(z)(z-w)-{2T\over \pi \sqrt{2m}}P(z)\,.
\label{gg}
\end{equation}
Such a function cannot have a zero, because if we assume that $R_w(\zeta)=0$, 
and $Q(\zeta)\ne 0$, then $\zeta$ is the partner of $w$ because 
\beno
\zeta=w+{2T\over \pi \sqrt{2m}} \frac{P(\zeta)}{Q(\zeta)}\,,
\eeno
and therefore 
$\zeta$ is a pole of $U(z)$, what is not possible because we assumed that
$Q(\zeta)\ne 0$. On the other side, had we assumed that $\zeta$ is a 
zero of $R_w$ for which $Q(\zeta)=0$, then 
(\ref{gg}) shows that also $P(\zeta)=0$, what is once again against our hypothesis that $P$
and $Q$ have no common zeroes. 

  As the polynomial function $R_w(z)$ has not zeroes, it should be a
 constant.  
 Now, if 
  we have two different zeros of $Q$, $w_1$ and $w_2$, the preceding argument
  shows the existence of two constants $c_1$ and $c_2$ such that 
\beno
{Q(z)} (z-w_1)-{2T\over \pi \sqrt{2m}}P(z)=c_1\,,\qquad {Q(z)}
(z-w_2)-{2T\over \pi \sqrt{2m}}P(z)=c_2\,, 
\eeno
from where we find that 
  \beno
w_1-w_2= \frac{c_2-c_1}{Q(x)}\,,
\eeno
 which implies that
$Q$ must be a  constant, reducing the problem to the previously considered
case of $U$ being  polynomial. Therefore  the only possible non-polynomial
 solution is the one
 given by
 a polynomial $P$ of degree two  and a polynomial $Q$ of degree one   with one
single real zero, which completes the proof of the claim.

Note that using translational symmetry we can fix the real pole at $x=0$,
(i.e. $d=0$) and  the classical motion can be then
 restricted to the open interval $(0,\infty)$.

Some other examples with a non-analytic behavior are the following.

{\bf Case C}. The half  harmonic
oscillator whose potential is
\be
U(x)=\cases{{\infty}& {if $x\leq 0$}\cr\cr
{\displaystyle{1\over 2}m \omega^2x^2}&{if $x\geq 0$}}
\label{lah}
\ee
is  only defined in half a line. However it does  not define a new family of
isochronous  potentials because it can be included in the Abel's family of 
the regular harmonic oscillator $U(x)= {{2}m \omega^2x^2}$. In fact, it is 
related to the oscillator  by the shear transformation defined by \cite{OO87}
\be
g(U)=-\frac{\sqrt U}{\sqrt{2m \omega^2}}\,.
\ee
Note that this half-harmonic system can be considered as the limit when
$\omega_1$ tends to infinity of the two half-oscillators system (\ref{twofreq}).
In fact, using the relation (\ref{semiT}) with $\omega_2=\omega$ and taking the
limit when $\omega_1$ tends to $\infty$ we obtain $\omega_0=2\, \omega$ 
and therefore the potential (\ref{lah}) is in the same equivalence class as
the harmonic oscillator given by $U(x)= {{2}m \omega^2x^2}$. 

Finally, as indicated before, this potential is obtained in  Case B for $\alpha=1$.

{\bf Case D}. To the same family belongs the potential \cite{{vesel},{NS79a},{NS79b},{NG81}}
\be
U(x)=\frac{2\, \alpha^2}{m \omega^2x^2}+{1\over 2}m \omega^2{x^2}-2\, \alpha=
{1\over 2}m \omega^2 \left(\frac{2\, \alpha}{m \omega^2x}-x\right)^2\ .
\ee
It is obvious that this potential is isochronous 
because in fact it is related with  the  half  harmonic
oscillator (\ref{lah}),
by means of a shear transformation
\be 
\displaystyle
g(U)=\sqrt{ U\over 2 m\omega^2}-\sqrt{4\, \alpha + U\over 2 m\omega^2}
\ee

\begin{figure}
\hskip3cm\includegraphics[width=.6\textwidth]{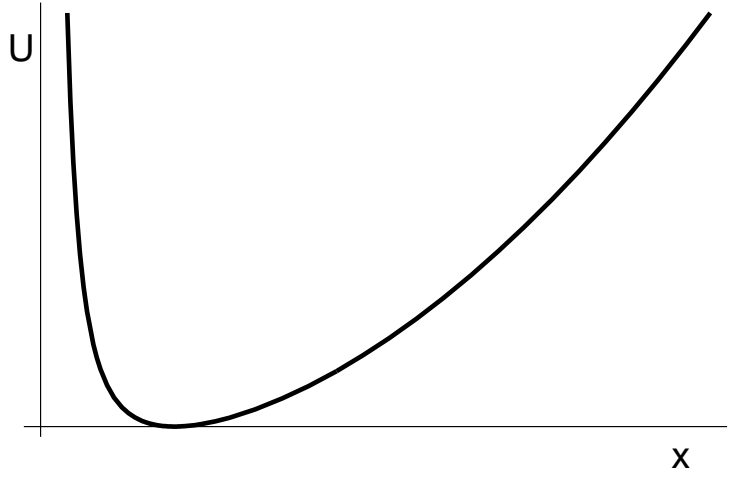}
\caption{Isochronous  potential $U(x)={1\over 2}m \omega^2 \left(\frac{2\, \alpha}{m \omega^2x}-x\right)^2$}
\end{figure}
and to the symmetric oscillator  
by the shear transformation \cite{OO87}
\be 
\displaystyle
g(U)=-\sqrt{4\, \alpha + U\over 2 m\omega^2}
\label{sh}
\ee

Note that according to Chalykh-Vesselov theorem \cite{vesel} (theorem 2),  
this potential and the harmonic oscillator potential are the only
rational isochronous potentials.

Another  characteristic case of the same family is the following isochronous
one:

{\bf Case E}. For  $g(U)=\alpha\, U$, then
\be
x_1(U)=- \sqrt{2U\over m\omega^2}+\, {2\alpha U\over m\omega^2}\,,\quad x_2(U)=\sqrt {2U\over m\omega^2}+ \,
{2\alpha U\over m\omega^2}\ ,
\ee
and therefore, for both values of $x$ we have
\be
\left(x-\, {2\alpha U\over m\omega^2}\right)^2={2U\over m\omega^2}
\ee
or in other form,
\be
\alpha^2\, {U^2 \over m^2 \omega^4}-\left(\alpha x+\frac 1{2}\right)\, {U\over m\omega^2}+\frac 14x^2=0\ ,
\ee
from which we obtain that \cite{PM00}, if $x\geq -1/(4\alpha)$,
\be
U(x)
={m\omega^2\over 2}\left[\frac x\alpha+\frac 1{2\alpha^2}-\frac 1{\alpha}
\sqrt{\frac x\alpha+\frac 1{4\alpha^2}}\right]\ .
\ee
\begin{figure}

\hskip3cm\includegraphics[width=.6\textwidth]{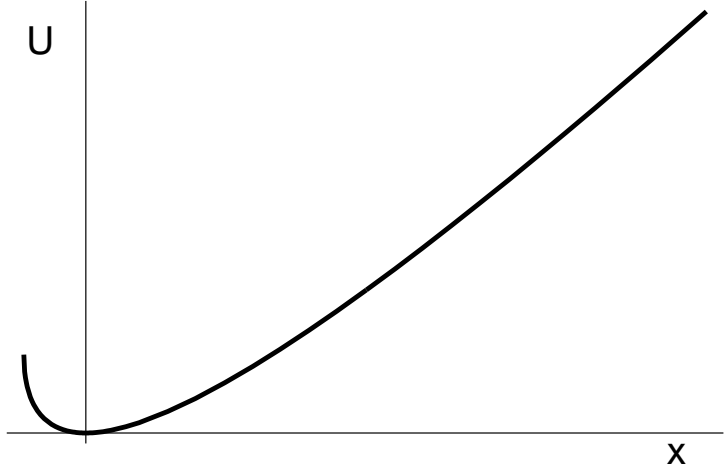}
\caption{Isochronous potential $U(x)={m\omega^2\over 2}\left[\frac x\alpha+\frac 1{2\alpha^2}-\frac 1{\alpha}
\sqrt{\frac x\alpha+\frac 1{4\alpha^2}}\right]$}
\end{figure}

Note that $U(0)=0$ and for small values of $x$,
\be
U(x)\approx {1\over 2}\,m\omega^2 x^2-\,{m\alpha\omega^2}\,  x^3+\cdots \ .
\ee

In this case, although the shear transformation of the oscillator is smooth the
final system is only defined on half a line.

{\bf Case F}. An archetypal case is the reduced Kepler problem (see
\cite{LL76}, Chapter III)
\be
U(x)=-{e^2\over x}+{l^2\over 2 m x^2} \quad {\rm for}\  x>0
\ee
whose period function for negative energies, is well known
\be
T(E)={\pi e^2}\sqrt{m \over 2 |E|^3}
\ee
and is a particular case of a more general family of  potentials
(see also
\cite{LL76}, Chapter III)
\be
U(x)=A\, |x|^n
\ee
with periods
\be
T(E)={2\over n}\sqrt{2\pi m \over E}\left({E \over  A}\right)^{1\over n}
{\Gamma\displaystyle\left({ 1\over n}\right)\over \Gamma\displaystyle\left({1\over 2}
  +{1\over n}\right)} \,.
\ee

{\bf Case G}. A very peculiar different example is the infinite wall, 
\be
U(x)=\cases{{0}&{if $x\in [0,\pi]$}\cr\cr
{\infty}&{if $x\notin [0,\pi]$}}
\ee
which is only shear equivalent to itself up to space translations.
In this case, the Abel inverse of the period function 
\be
T(E)=\pi\, \sqrt{2m \over {E}}
\ee
is uniquely defined up to a shift by a real constant  $a$.

{\bf Case H}. A similar potential with the same quantum energy spectrum
\be 
U(x)=\frac 1m\left({1\over \sin^2(x)}-\frac 12\right)
\ee
has a much larger degeneracy \cite{RWR}. 

The last two cases show that the orbits of Abel's shear
transformations are not of the same type.
\begin{figure}

\hskip3cm\includegraphics[width=.6\textwidth]{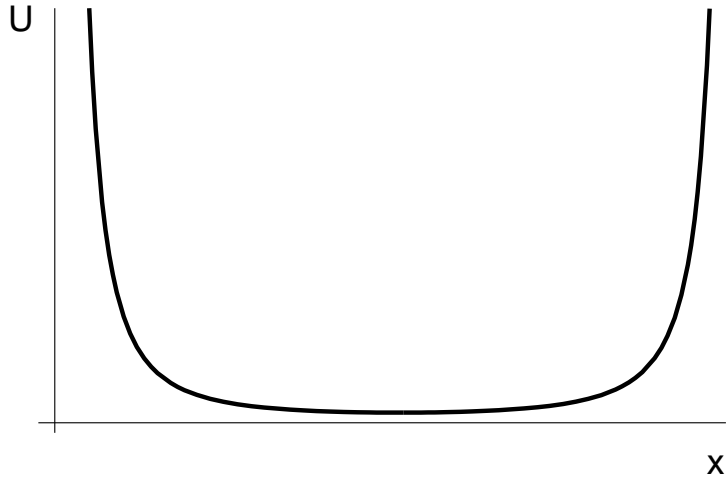}
\caption{Smooth well potential $U(x)={1\over \sin^2(x)}$ with the same energy
spectrum that the infinity square well.}
\end{figure}

For even potentials there is a special case of  shear transformation
which preserves the periods. It is given by a composition of an 
inversion with two translation
transformations.

The transformation of the complex plane called Joukowski transformation, 
 defined by  $J_\lambda(z)=z+(\lambda/ z)$, with $\lambda\in {\R}$,
plays a relevant in  aerodynamics applications. We consider here an
 analogous map of the real line completed with the two points at the
infinity:
\beno
J_g(x)=\frac x2-\frac {2\,g^2}x\,,\qquad \,.
\eeno
We also consider the involution of $\bar{\mathbb{R}}$, $i_g:\bar{\mathbb{R}}\to
\bar{\mathbb{R}}$, given by
\be
i_g(x)=- \frac{4g^2}{x} \,.
\ee

Note that $J_g(0+)=-\infty$ and $J_g(\pm \infty)=\pm \infty$ and the important property
$J_g\circ i_g=J_g$. Consequently, the points $x$ and $-4g^2/x$ have the same
image. Moreover, only these two points have the same image, because if 
$x/2-2g^2/x=y$, then $x^2-2xy-4g^2=0$, and therefore the two roots are given by
\beno
x_{\pm}(y)=y\pm \sqrt{y^2+4g^2}\,,
\eeno
i.e. $x_+(y)>0$, $x_-(y)<0$ and $x_+(y)\, x_-(y)=-4g^2$.

 We can use the properties of these transformations $J_g$ and $i_g$ to prove:

{\bf Theorem 3.}
If $ U(x)$ is a bounded below even  convex potential with $\lim_{x\to \infty }
U(x)=\infty$, then for any 
real number $g$ the potential $U_g$ given by 
\beno
U_g(x)=U(J_g(x))=U\left(\frac x2-\frac{2\, g^2}x\right)
\eeno
   is
isoperiodic with $U(x)$. 

{\it Proof:}  First notice that $U_g$ is invariant under the 
transformation $i_g$, because $U_g(i_g(x))=U(J_g(i_g(x)))=U(J_g(x))=U_g(x)$. 
The parity symmetry of the function $U$ implies that  
$U_g(x)=U_g(4 g^2/x)$. 

On the other side, as the function $U$ is a bounded below even  convex potential
the minimum of the potential is at the origin and  we can consider without any
restriction that the minimum value of
is $U(0)=0$.  

If $U_g(x_1)=U_g(x_2)$, then $U(J_g(x_1))=U(J_g(x_2))$, and therefore, 
given  an arbitrary  positive energy value
$E>0$ there will be two real numbers $x_-(E)<0$ and $x_-(E)>0$ such that 
$-x_-(E)=x_+(E)$ and $U(x_\pm(E))=E$. Consequently, using the definition of
the new potential function $U_g$, there will exist four points, to be denoted 
$x_{g_1}^-,x_{g_2}^-,x_{g_1}^+,x_{g_2}^+$ such that $U_g(x_{g_i}^\pm)=E$. They are respectively given by 
\begin{eqnarray}
x_{g_1}^-&=&-x_+(E)-\sqrt{(x_+(E))^2+g^2}\,,\qquad 
x_{g_2}^-=x_+(E)-\sqrt{(x_+(E))^2+g^2}\,,\cr
x_{g_1}^+&=&-x_+(E)+\sqrt{(x_+(E))^2+g^2}\,,\qquad
x_{g_2}^+=x_+(E)+\sqrt{(x_+(E))^2+g^2}\,.\nonumber
\end{eqnarray}   

 The  span between the
two $U_g$--equipotential values 
$x_{g_1}^-$ and 
$x_{g_2}^-$, and same for $x_{g_1}^+$ and $x_{g_2}^+$, 
is $2x_+(E)$, and it coincides  with the span between the
corresponding $U$-equipotential values
 of the parity symmetric potential $U$. Consequently, the
potentials $U_g$ and $U$ are shear related and isoperiodic\footnote{Strictly speaking 
$U_g$ has two branches, one in each half--line of positive/negative values of
$x \in \IR$. Therefore, there is a degeneracy of trajectories which is not
present in the  convex potential $U$ which has only one branch. }.

In the case of isochronous potentials this connection between pairs of
potentials provides us with the only solutions to isochronous rational potentials
in terms of the harmonic oscillator and the isotonic potential (see theorem 2).
This result can be further generalized. Indeed it can be shown that in the rational 
case these two families of potentials are the only ones which are isoperiodic 
not only for the isochronous periods but for any frequency-energy spectral 
distribution associated to a rational potential.

{\bf Theorem 4.}  Any non-trivial rational  potential $U_\ast$
   which is isoperiodic to a given
 even convex polynomial potential $U$ is either of the form
$U_c=U(x+c)$ or  $U^c_g(x)=U((x-c)/2-2g^2/(x-c))$, for any value of $g$.

{\it Proof}: If there is a  rational potential solution of 
\be
{U_\ast}(x)= U_\ast\left(x+W_U(U_\ast(x))\right).
\label{rrg}
\ee
the function $W_U(U_\ast (x))$ has to be the irreducible ratio of two polynomials 
$P(x)$ and $Q(x)$, i.e.
\beno
W_{U}(U_\ast(x))=\frac{P(x)}{Q(x)}.
\eeno
The stability of the
leading term under the non-linear constraint (\ref{dd}) requires that
the degree of  $P$ has to be one unit larger than that of $Q$. 
On the other hand by construction the transformation $K$ defined by 
\beno
K(x)=x+\frac{P(x)}{Q(x)}
\eeno
has to be invertible and involutive, i.e. $K\circ K=\Id$.
It is easy to show 
 that the only rational solutions satisfying this
requirement
are 
\be
K_c(x)=-x+c\qquad K^g_c(x)=c -\frac{4 g^2}{x-c}
\label{involution},
\ee 
which correspond to the kind of  transformations, {\it translations, reflections and inversions}, described previously
in this section. 

Indeed, if $K_c$ is polynomial the asymptotic analysis at $x\sim \infty$ requires that the leading term
$K_c\sim a_n x^n$ satisfies $K_c(K_c (x))\sim a_n^{n+1} x^{n^2}\simeq x$ and 
thus $n=1$ and $a_n^2=1$.  The only
non-trivial solutions of these requirements are the translations/reflections of (\ref{involution}).
This  regular  type of solutions keeps the polynomial
character of the  potential and  simply involve   a reflection and
a translation of the polynomial.

If $K_c$  is rational it can have  poles in the complex plane. Notice that because of the rational character
of the transformation $K_c$ the involutive property can be extended to the whole complex plane.
If  $K_c$ is not a pure polynomial it   has to have a  pole at a point $c\neq \infty$ which is the image of $x=\infty$.
and  can be rewritten in the form
\be
K_c(z)=c-\frac{P_0(z)}{(z-c) Q_0(z)}.
\ee
Since
\be
\displaystyle
K_c\circ K_c(z)=c-\frac{P_0(K_c(z)) {Q_0(z)} (z-c) }{ {P_0(z)} Q_0(K_c(z))}=z
\ee
we have that
\be
1=\frac{P_0(K_c(z))Q_0(z)}{Q_0(K_c(z))P_0(z)}.
\ee
It is easy to show that the only solution is 
$Q_0(z)=P_0(z)={\mathrm {cte}}$ and the pole has to be a real pole, i.e $c^\ast =c$. 
The absence of other poles is excluded by the involutive character of the transformation, i.e.
only one point in the complex plane can be involutively mapped into  $z=\infty$.

The second kind of solutions of (\ref{involution}) is more subtle and implies that $P=4 g^2 -(x-c)^2$
and $Q=x-c$, which means that $W_U(U_\ast (x))$ and therefore $U_\ast $ develops a single
pole singularity at $x=c$. In this case $U_\ast$ is symmetric under inversion
transformations,
\be
U_\ast (x)=U_\ast\left(c+\frac{4 g^2}{x-c}\right)
\label{inv}
\ee
the same symmetry properties that the potential $U^c_g$ satisfies.
Now, since  $U$ is convex even potential its minimum is attained at $x=0$. The minimum of
$U_\ast$ for $x>c$ is at $x=c+{2}g$ and 
since $U_\ast$ is isoperiodic to $U$  the values of the two potential at both minima have to be identical,
i.e. $U_\ast({2}g)=U(0)$.
By theorem 3 the potential $U^c_g$ is also isoperiodic to $U$,  has the same symmetry
that $U_\ast$ under inversion transformations (\ref{inv})  and verifies that $U^c_g({2}g)=U(0)$ .

Now, since $U^c_g$ and $U_\ast$ are isoperiodic both must
 attain the same values at $x$ and $c+{4 g^2}/{x-c}$, which implies that
 $U_\ast=U^c_g$ and proves the theorem. $\Box$

In particular, the only singular rational potentials which are isoperiodic to
$U(x)=x^2$ and are  singular at $x=0$ are those of the form $U(x)=(x/2-2g^2/x)^2$

\section{Scale transformations and Isoperiodicity}

There is another kind of transformations which also preserves isoperiodicity.
They are connected with  space-time scale transformations.

The time-evolution of   one-dimensional  systems is described in
terms of the 
potential function $U(x)$, i.e. $\ddot x=-\partial U/\partial x$. If we
introduce a  change of space-time coordinates defined by 
\be
x=\beta\, \tilde x\,,\qquad  t=\sqrt \gamma\, \tilde t\,,
\label{scale}
\ee
where $\beta$ and $\gamma$ are positive real numbers, then 
the equation of motion becomes
\beno
\frac{\beta}{\gamma}\frac{d^2\tilde x}{d\tilde  t^2}=-
\frac{1}{\beta}\left(\pd{U}{\tilde x}\right)(\beta\, \tilde x)
\eeno
and therefore, if we define 
\begin{equation}
\wt U(x)=\left(\frac{\gamma}{\beta^2}\right)\, U(\beta\, x)\,,\label{Utilde}
\end{equation}
we find that the equation of motion reads
\beno
\frac{d^2\tilde x}{d\tilde  t^2}=-\pd{\widetilde U(\tilde x)}{\tilde x}\,.
\eeno
This invariance of the equation of motion is a consequence of the transformation
of the action:
\beno
S(x)=\int dt \left(\frac{1}{2} \dot{x}^2 - U(x)\right)\,,\qquad \tilde{S}=\frac{\gamma}{\beta^2}S\,.
\eeno
This suggests to study the relation between systems described by potentials
$U$ and $\wt U$ related as in equation (\ref{Utilde}). 

We introduce next a generalization of a property studied by 
Dorignac \cite{JD05}.

Let $\varphi (\zeta)$ an arbitrary function and define 
\beno
I_\varphi(E)=\int_{x_-(E)}^{x_+(E)} \varphi(E-U(x))\, dx\ .
\eeno

If for any pair of  real numbers $\beta,\gamma\in {\af R}$ 
we define a new potential given by 
\beno
\wt U(x)=\left(\frac{\gamma}{\beta^2}\right)\, U(\beta\, x)\,,
\eeno
then 
\beno
\wt x_\pm (E)=\frac 1 \beta\, x_\pm\left(\frac{\beta^2\, E}{\gamma^2}\right)\,,
\eeno
and consequently
\beno
\wt I_\varphi(E)=\int_{\wt x_-(E)}^{\wt x_+(E)} \varphi(E-U(x))\, dx\ .
\eeno

Therefore,  
\beno
\wt I_\varphi(E)=\int_{(1/\beta)x -(\beta^2\,
  E/\gamma^2)}^{(1/\beta)x+(\beta^2\,
  E/\gamma^2)}\varphi\left(E-\frac{\gamma^2}{\beta^2}\, U(\beta\, x)\right)\,dx=
\frac 1\beta\int_{x_-(\beta^2\,
  E/\gamma^2)}^{x_+(\beta^2\,
  E/\gamma^2)}\varphi\left(E-\frac{\gamma^2}{\beta^2}\, U(y)\right)\, dy\,,
\eeno
 and defining  
\beno
\widetilde E=\frac{\beta^2\, E}{\gamma^2}
\eeno
the equation can be rewritten as
\beno
I_\varphi(E)=\frac 1\beta\int_{x_-(\wt E)}^{x_+(\wt E)}
\varphi((\gamma^2/\beta^2)\,(\wt E-U(y)) \, dy\,.
\eeno
 If the function $\varphi$ is homogeneous of degree  $p$,
\beno
\wt I_\varphi(E)=\frac {\gamma^{2p}}{\beta^{2p+1}}\, I_\varphi(\wt E)\,.
\eeno

When computing the period of an oscillating motion we find a function as
$I_\varphi$ with $\varphi $ a function proportional to 
$\varphi_P(\zeta)=\zeta^{-1/2}$ and when computing the action we arrive to
 a  function $\varphi_a(\zeta)=\zeta^{1/2}$. Therefore,
\beno
\wt I_{\varphi_P}(E)=\frac 1{\gamma}\, I_{\varphi_P}(\wt E)\,,\qquad 
\wt I_{\varphi_a}(E)=\frac {\gamma}{\beta^2}\, I_{\varphi_a}(\wt E)\,.
\eeno

As a consequence If $U(x)$ is an isochronous  potential with period $P$,
then ${\widetilde U}$ is isochronous too and its period is 
$\wt P=P/\gamma$. In
particular, for $\gamma=1$ we obtain that if $U(x)$ is an isochronous
potential with period $P$, then  $\wt U(x)=\beta^{-2}\, U(\beta\, x)$ is
isochronous too with the same frequency \cite{JD05}.

In particular, for $\gamma = 1$ we see that if  $U (x)$ is an isochronous 
potential, then $\wt U(x)=\beta^{-2}U(\beta\, x)$ is isochronous too and with
the same period. On the contrary, the action for this potential $\wt U$ 
is not the same as for $U$, and if the spectrum of the first one is
equispaced, it will not be true for the new potential.

In many cases the above scale transformation  can be shown to be equivalent
to a shear transformation, e.g. in some of the examples of the previous section
are related by scale transformations. However, the scale transformation
has a different nature. It establishes an equivalence relation among
isoperiodic potentials but does not preserve the energy levels unlike the shear
transformation. For such a reason one does not expect that quantization will
preserve the equivalence at the spectral level.

In physical terms the scale transformation has a energy cost whereas the
classical shear transformation is energy preserving. An interesting question is
to  know whether or not the quantization prescription preserves this 
classical property. This will be the subject of next section.

\section{Isoperiodicity and the quantum isospectrality}

It is clear from the analysis of Section 2 that two   potentials $U_1$ and
$U_2$  related by a
shear transformation $g:\IR\to \IR^+$ define similar period functions for periodic orbits.
In fact, not only the periods given by (\ref{period}) are identical
for $U_1$ and $U_2$ but also any
integral between the same limits of the form
\be
T_f(E)=\int_{x_m(E)}^{x_M(E)}f\left({{E-U(x)}}\right){dx}
\label{area}
\ee
is the same for both potentials.
The proof is simple because the integral (\ref{area}) can be splitted
as 
\beno
T_f(E)=\sum_{i=1}^NT_f^i(E)
\eeno
in terms of the integrals
\beno
T_f^i(E)=\int_{x_i(E)}^{x_{i+1}(E)}f\left({{E-U(x)}}\right){dx}
\eeno
where  $x_i(E)$, $i=0,1,\ldots,N$,  is the monotone sequence of points 
whose initial and final points are 
$x_0=x_m$, $x_N=x_M$, i.e.  they 
 coincide with the turning points of the classical
trajectory,
 and the remaining  points  $x_i$, for $i=1,2,\dots, N-1$,  are defined
by  the values $x_i\in [x_m,x_M]$  for  which there is a stationary point
 $x_i^\ast\in[x_m,x_M] $ of the potential  $U'(x_i^\ast)=0$ with the same potential level
$U(x_i^\ast)=U(x_i)$.
In each interval $ [x_i,x_{i+1}]$ the potential function  $U(x)$ is
invertible and the inverse function $x_i(U)$ is uniquely defined.
Thus,
\beno
T_f^i(E)=\displaystyle \int_{U(x_i)}^{U(x_{i+1})}\, dU\, f(E-U)\,  x'_i(U)\ .
\eeno
Now, by construction,  for each interval $[x_i,x_{i+1}]$ there is another
one $[x_{i'},x_{i'+1}]$ such that the sum of the contributions to the
the integral (\ref{area}) 
\begin{eqnarray}
T^i_f(E)+T^{i'}_f(E)&=\displaystyle \int_{U(x_i)}^{U(x_{i+1})}\, dU\, f(E-U)\,
|x'_i(U)-x'_{i'}(U)|
\label{equiv}
\end{eqnarray}
becomes the same for the two potentials $U_1$ and $U_2$.  
In fact, the integrand  and the integral limits in (\ref{equiv}) are identical for 
any couple  $U_1$ and $U_2$ of shear equivalent potentials. In particular, this shows that
  $T_f$ is also the same for all potentials related by a shear transformation and that their
first semiclassical quantum corrections to the energy levels, which is given by $T_f(E)$ with $f(x)=\sqrt{x}$,
is also the same.

From the discussion of  the previous section it follows that scale
transformations (\ref{scale}) also preserve semiclassical corrections to 
energy levels if $p=1/2$ and $\gamma=\beta$. However in such a case the
classical system is not isochronous.

However the higher order corrections might break the equivalence at
the quantum level. The case B considered in the previous section is
the simplest counterexample. The energy levels $E^B_n$  \cite{SS89}
differ from those of the isoperiodic harmonic oscillator
\be
E^A_n= {\hbar\omega}\left(n+\frac12\right)
\ee
by terms which start at first order in perturbation theory  for small values of 
the anharmonicity parameter $\alpha<<1$

\begin{eqnarray}
\begin{array}{lll}
E^B_n-E^A_n \!\!&=&\!\!\displaystyle \frac{\hbar\,\omega_0\,\alpha^2(3-\alpha^2)}{4 (1-\alpha^2)^2}
\left(n+\frac12\right)  + \frac{\hbar\,\omega_0\,\alpha^2}{ (1-\alpha^2)^2}
\frac{2n + 1}{8}  \left({(2n + 1)} \psi\left(-(-1)^n \frac{n}{2} + \frac12\right)\right.\cr 
\noalign{\vskip 12pt}
 &  & \displaystyle \left. 
-  {(1 + 2n)}
\psi\left((-1)^n \frac{1+n}{2} + \frac12\right) -1\right)
+ O\left(\frac{\alpha^4}{(1-\alpha^2)^4}\right)\crcr
\noalign{\vskip 12pt}
& =&\displaystyle
{\hbar\,\omega_0\,\alpha^2}
\frac{2n + 1}{8}  \left(2- {(1 + 2n)}
\psi\left((-1)^n \frac{1+n}{2} + \frac12\right) \right.\cr
\noalign{\vskip 12pt}
& &\displaystyle \left.   +{(2n + 1)} \psi\left(-(-1)^n \frac{n}{2} + \frac12\right)\right)
+ O(\alpha^4)
\nonumber
\end{array}
\end{eqnarray}
where $\psi(x)$ is the logarithmic derivative of the Euler gamma function $\Gamma(x)$. The  above perturbative
expression for $E^B_n$ agrees to order $\alpha^2$  with the asymptotic behavior derived from
 the exact spectral equation \cite{SS89}
\be
\displaystyle
\frac{\displaystyle\Gamma\Bigl( \frac34 -\frac12(1+\alpha) E^B_n\Bigr)}{\displaystyle\Gamma\Bigl( \frac34 -
\frac12(1-\alpha) E^B_n\Bigr)}+
\displaystyle
\frac{\displaystyle \sqrt{1+\alpha}\,\,\,\Gamma\left( \frac14 -\frac12 (1+\alpha) E^B_n\right)}{\displaystyle\sqrt{1-\alpha}\,\,\,
\displaystyle\Gamma\left( \frac14 -\frac12(1-\alpha) E^B_n\right)}=0.
\ee

One particular case where the quantum energy levels remain equal
is when the shearing function $g$ is constant, $g={\mathrm{const.}}=g_0$. In such a case both potentials are
related by a simple translation $U_2(x)=U_1(x-{g_0})$, which obviously
does not change the quantum spectrum of the Hamiltonian.
Further non-trivial examples can be obtained by means of Darboux
method.

\section{Shear deformation and Darboux transform}

The quantum spectrum is also the same, up to a shift,  for
two potentials related by a shear transformation  when they
can be written in the form
\be
U_1(x) =\frac{\hbar^2}{2m }\left(W(x)^2-W'(x)\right)-a_1\,;\quad U_2(x) =\frac{\hbar^2}{2m }\left(W(x)^2+W'(x)\right)-a_2\,,
\label{darboux}
\ee
in terms of a common superpotential $W(x)$ with 
$\lim_{x\to\pm \infty}W(x)=+\infty$ and two constants $a_1$ and $a_2$. 
Potentials of such a type are not only related by a classical 
shear transformation but they are also related by
a quantum Darboux transformation \cite{Darboux} which guarantees that
 the corresponding spectra of the Hamiltonians
\beno
H_i=\frac{p^2}{2m}+U_i\quad i=1,2
\eeno
are almost identical\footnote{The asymptotic behavior of $W$ guarantees that
the ground states of the two systems are in one to one correspondence and none
has zero energy  for $a_1=a_2=0$.}. 

The case  D of Section 3 is also an example of such a type.
In fact, choosing 
\be
W(x)=\frac{1}{x}+x\ee
and $a_1=1+2\sqrt{2},a_2=3
, \hbar^2=2 m$
we have the two potentials of case D 
\be
U_1(x)=\frac{2}{x^2}+x^2-2\sqrt{2};\qquad
U_2(x)=x^2
\label{isotonic}
\ee
provided we fix $m\,\omega^2=2$ and $\alpha=\sqrt{2}$ for simplicity.
It is also clear from the discussion of previous section
that both potentials are related by the shear transformation  
\be 
\displaystyle
g(U)=+\frac{\sqrt{U}}2 + \sqrt{\frac {4\, \sqrt{2} + U} 2}\,.
\label{sh2}
\ee

More generally, for any choice of  the superpotential 
$W$ with parity invariance $W(x)=W(-x)$, i.e. $W$ is of the form
$W(x)=K(x^2)$, it can be shown that the corresponding potentials $U_1, U_2$
are related by a parity symmetry $U_1(x)=U_2(-x)$. If $U_1$  and $U_2$ are convex
functions  they are obviously related by the shear transformation.

However, it should also emphasized that not any pair of 
potentials of the form (\ref{darboux}) related by a Darboux transform
are  necessarily related by a classical shear transformation. A simple 
counterexample is given by $W(x)=x^4-x$. In that case one gets
the potentials 
\be
U_1(x)=x^8 - 2x^5- 4 x^3+ x^2+1  
\label{uno}
\ee
and 
\be
U_2(x)=x^8 - 2x^5+ 4 x^3+ x^2-1, 
\label{dos}
\ee
 respectively. It is obvious from the Figure 7
that both potentials are not shear related.
\begin{figure}
\includegraphics[width=.5\textwidth]{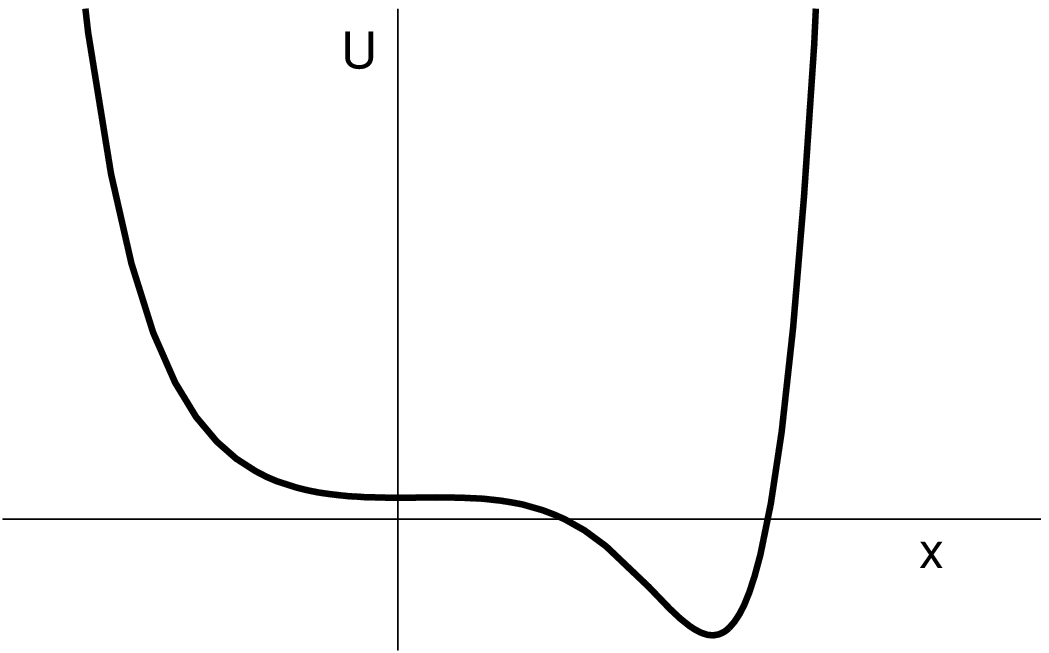}
\vskip-4.54cm \hskip8cm
 \includegraphics[width=.5\textwidth]{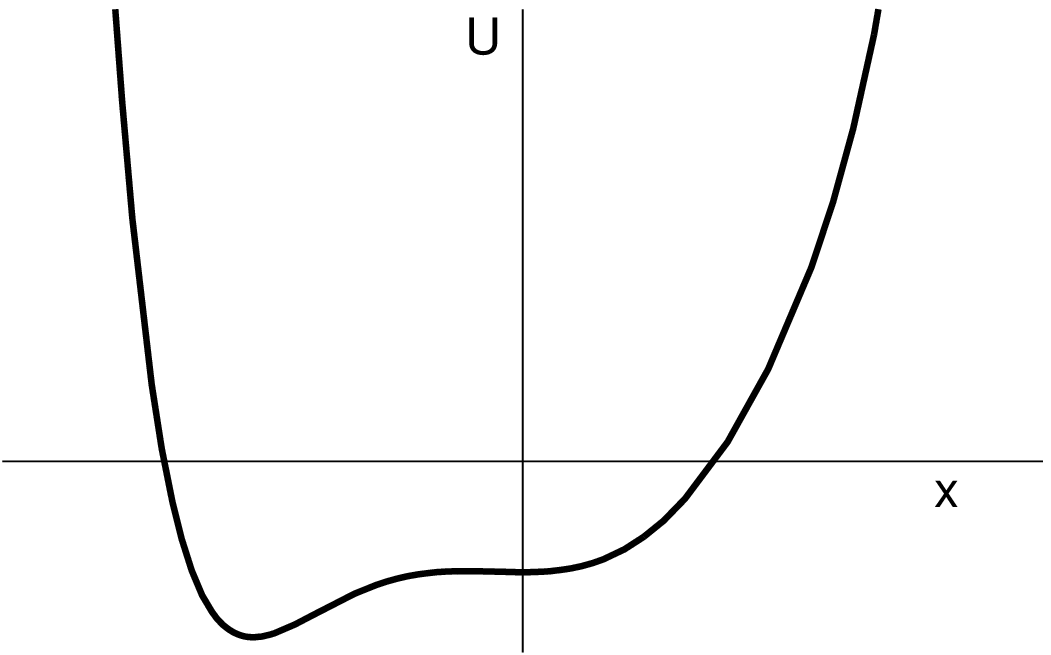}

\caption{Pair of isospectral potentials (\ref{uno}) and  (\ref{dos}) which are not isoperiodic}
\end{figure}

This illustrates that the generalization of theorems 1 and 2 to the quantum
case es more sophisticated.

There are further examples of quantum isospectral systems which are not
classically isoperiodic. A very interesting case is the following 
\cite{AM80,{Gelfand-Levitan},{mactrub},{PZ98},{JoK52}}.
Let us consider a standard quantum oscillator ($m=\omega=1$)
with Hamiltonian 
\be
H_0=\frac 12(p^2+x^2)\ .
\ee

We know that the eigenvalues and eigenstates of such operator are given by 
\be
H_0\,\varphi_n(x)=E_n\,\varphi_n(x)\ ,
\ee
with 
\be
 E_n=\left(n+\frac 12\right)\,,
\quad n=0,1,2,\ldots\,,
\ee
 and 
\be
\varphi_n(x)=(\sqrt\pi\, 2^n\, n!)^{-1/2}\ {\cal H}_n(x)\, \exp\left(-\frac{x^2}2\right)\ ,
\ee
where ${\cal H}_n(x)$ denotes the Hermite polynomial.

Let us now consider the system with the same spectrum, except the lowest eigenvalue $E_0=1/2$:
\be
E_n=\left(n+\frac 12\right)\,,
\quad n=1,2,\ldots\,.
\ee
For this we perform a similarity transformation  which maps $\varphi_0(x)$ 
 into $\widetilde \varphi_0(x)$ by the formula \footnote{A similar transformation based
on modding out by any eigenstate $\varphi_n$ can be formally achieved but because of
the existence of nodes in the wave function $\varphi_n$ the induced potential is not
defined on the whole real line $\IR$ \cite{JoK52}}
\be
\widetilde \varphi_0(x)=\frac{\varphi_0(x)}{\Phi(x)}
\ee
where 
\be
\Phi(x)=\int_x^\infty \varphi_0^2(\xi)\,d\xi=\frac 1{\sqrt \pi}
\int_x^\infty {\rm e}^{-\xi^2}\,d\xi\ ,
\ee
for which 
\be
\Phi'(x)=-\varphi_0^2(x)=-\frac 1{\sqrt \pi}\, e^{-x^2}\ .
\ee
\begin{figure}
\hskip3cm\includegraphics[width=.5\textwidth]{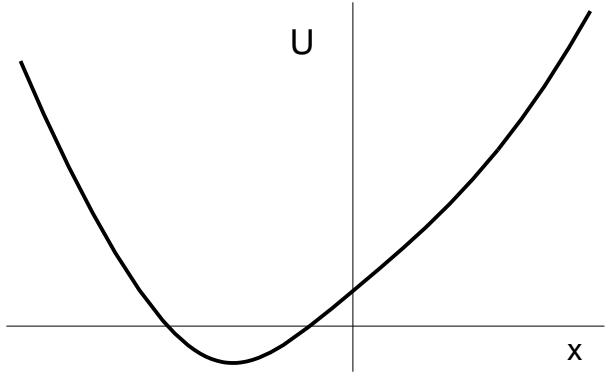}
\caption{Potential $U(x)=\frac12 x^2+4\, \chi(x) (\chi(x)-x)$ with equally spaced spectrum }
\end{figure}
Note that 
\be
\Phi(x)=\cases{ {1}& {{ at}\ $x\to-\infty$}\cr\cr
{\displaystyle \frac 1{2\,x}\,\varphi_0^2(x)} &{{ at}\ $x\to\infty$}}
\ee
and so,
\be
\widetilde\varphi_0(x)=\cases{ {\varphi_0(x)}& {at\ $x\to-\infty$}\cr\cr{\displaystyle
{\frac {2\,x}{\varphi_0(x)}}} &{at \ $x\to\infty$}},
\ee
and then $\int_{-\infty}^\infty |
\widetilde\varphi_0(x)|^2\, dx=\infty$.
Hence, 
\be
H_0\,\widetilde \varphi_0(x)=\frac 12\, \widetilde \varphi_0(x)\, ,\qquad{\rm at}\ x\to\pm\infty\, .
\ee

The function  $ \widetilde \varphi_0(x)$ is the solution of the equation
\be
H\widetilde \varphi_0=\frac 12\, \widetilde \varphi_0\ ,
\ee
where
\be
H=\frac 12\, p^2+U(x)
\ee
and 
\be
U(x)=U_0(x)+U_1(x)
\ee
with 
\be
U_0(x)=\frac 12 \, x^2\,,\qquad U_1(x)=-2\,\frac{d^2}{dx^2}
\log[\,{\rm erfc\,}(x)]=-4\, \chi(x) (\chi(x)-x)\ ,
\ee
where the functions ${\rm erfc\,}(x)$ and $\chi(x)$   are
\be
{\rm erfc\,}(x)=\frac 2{\sqrt \pi}\int_x^\infty \exp \left(-\xi^2\right)\, d\xi\ ,
\ee
which satisfies 
\be
{\rm erfc\,}(x)=\cases{{ 2}&{ $ x\to -\infty$}\cr\cr 
{1}&{$x=0$}\cr\cr
{\displaystyle\frac 1{\sqrt \pi\, x}\,{\rm e}^{-x^2}}& {$x\to\infty$}}
\ee
and 
\be
\chi(x)= (\sqrt\pi \
{\rm erfc\,}(x))^{-1}\,\exp(-x^2)\ ;\qquad \chi(x)\approx x\quad  {\rm at }\ x\to\infty\,.
\ee
Notice that the new potential $U(x)=U_0(x)+U_1(x)$ is neither shear equivalent
to the harmonic oscillator $\frac12 x^2$ nor  isochronous.

It can be shown  that $H_0$ and $H-{\uno}$ have the same spectra. Indeed, 
\be
H \widetilde \varphi_n(x)=E_n\,  \widetilde \varphi_n(x)\ ,\qquad n=1,2,\ldots
\ee
where 
\be
 \widetilde \varphi_n(x)= \varphi_n(x)-\sqrt{\frac2n}\ \chi(x)\, \varphi_{n-1}(x)\ ,
\ee
and the functions $\widetilde \varphi_n(x)$ are normalized:
\be
\int_{-\infty}^\infty |\widetilde  \varphi_n(x)|^2\, dx=1
\ee
and satisfy the completeness condition 
\be
\sum_{n=1}^\infty  \widetilde  \varphi_n(x)\,\widetilde  \varphi_n(y)=\delta(x-y)\ .
\ee
i.e  the Hamiltonians $H_0$ and $H-\uno$ are isospectral.
The peculiarity of this case is that the two isospectral potentials are
neither classically isoperiodic nor related by a Darboux transformation. 

Finally, it is also remarkable that the two families of isospectral rational potentials
connected by  Joukowski transformations as in Theorem 4 are in general not isospectral.
Only the case of isochronous potential the  half harmonic oscillator and the  potentials
(\ref{isotonic})  present  the same quantum spectrum.

\bigskip
{\parindent 0cm{\Large \bf Acknowledgments.}}

\bigskip
We thank F. Falceto, C. Farina, M. Ra\~nada, 
A. Segu\'\i{} and G. Sierra for discussions.
Support of projects  BFM-2003-02532, FPA-2003-02948, DGA2006 Grupo de Altas
Energ\'\i as and SAB2003-0256 is acknowledged.



\begin{thebibliography}{99}
\bibitem{aim}M. Asorey, A. Ibort and G. Marmo, {\it ``Global Theory of Quantum
    Boundary Conditions and Topology Change''}, Int. J. Mod. Phys. {\bf A 20}
  (2005) 1001--1025.
\bibitem{aim2}M. Asorey, A. Ibort and G. Marmo, {\it ``Boundary Conditions and
    Path Integral''}, Proceedings
of A. Galindo Festschrift, Eds. Alvarez-Estrada {\sl  et al}, Madrid
(2004) 165--173.
\bibitem{Ab26}{N.H. Abel},
{\it ``Aufl\"osung einer mechanischen Aufgabe''},
{J. Reine Angew. Math.} {\bf 1}  (1826) 153--57.
\bibitem{Steiner} R. Subramanian and K.V. Bhagwat, 
{\it ``A lower bound for ground-state energy by Steiner symmetrisation of 
the potential''}, 
J. Phys. A: Math. Gen.  {\bf 20}, 69--78 (1987)
\bibitem{BMK03} S. Bolotin and R.S. MacKay, {\it ``Isochronous potentials''}, 
in: {\sl
    Localization and energy transfer in nonlinear systems}, p. 217--224, eds. L. V\'azquez,
  R.S. MacKay
and M.P Zorzano, World Sci. (2003).
\bibitem{calog}F. Calogero, {\it ``Two new classes of isochronous Hamiltonian systems''},
J. Nonlin. Math. Phys. {\bf 11} (2004) 208--222.
\bibitem{huygens} Ch. Huygens, {\sl ``Horologium Oscillatorium''}, Paris (1673)
\bibitem{vesel} O. A. Chalykh and A. P. Veselov, {\it ``A remark
on rational isochronous potentials''}, J. Nonlin. Math. Phys. {\bf 12}  
Suppl. 1 (2005) 179--183.
\bibitem{EKK97}V.M. Eleonskii, V.G. Korolev and N.E. Kulagin, {\it ``On a classical analog of the
  isospectral Schr\"odinger problem''}, JETP Lett. {\bf 65}  (1997) 889--93.
\bibitem{JD05}J. Dorignac, {\it ``On the quantum spectrum of isochronous
    potentials''}, 
J. Phys. A:Math. Gen. {\bf 38} (2005) 6183--210.
\bibitem{LL76}{L.D. Landau and E.M. Lifshitz}, 
{\sl ``Mechanics''}, Pergamon Press, London  (1981).
\bibitem{BFK1} B.F. Kimball, {\it Three theorems applicable to vibration
    theory},
Bull. Amer. Math. Soc. {\bf 38}, 718--23 (1933).
\bibitem{BFK2} B.F. Kimball, {\it Note on a previous paper}, 
Bull. Amer. Math. Soc.
{\bf 39}, 386 (1933)
\bibitem{AHC} A.H. Carter, {\it ``A class of inverse problem in physics''}, 
Amer. J. Phys. {\bf 68} (2000) 698--703. 
\bibitem{Appell} P. Appell, {\sl ``Trait\'e de mechanique rationalle''},
Vol  {\bf 1}, Gauthiers-Villars, Paris (1902).
\bibitem{SS89}F.H. Stillinger and D.K. Stillinger, {\it ``Pseudoharmonic oscillators and Inadequacy
of Semiclassical Quantization''}, J. Phys. Chem. {\bf 93}  (1989) 6890--92.
\bibitem{OO87}E.T. Osypowski and M.G. Olsson, {\it ``Isynchronous motion in
    classical mechanics''}, Amer. J. Phys. {\bf 55} (1987) 720--25. 
\bibitem{PM00} P. Mohazzabi,
{\it ``On classical and quantum harmonic potentials''},
Can. J. Phys. {\bf 78} (10) (2000) 937--946.  
\bibitem{GH}G. Ghosh and R. W. Hasse, {\it ``Inequivalence of the classes of
    quantum and classical harmonic potentials: Proof by example''},
  Phys. Rev. {\bf D 24}  (1981) 1027--29.
\bibitem{NS79a} M.M. Nieto and  L.M. Simmons, {\it``Coherent states
 for general
  potentials. I. Formalism''}, Phys. Rev. {\bf D 20} (1979) 1321-31 
\bibitem{NS79b} M.M. Nieto and L.M. Simmons,  {\it``Coherent states
 for general
  potentials. II. Confining one-dimensional examples''}, Phys. Rev. {\bf D 20}, 1332--41 (1979)
\bibitem{NG81} M.M. Nieto and V.P. Gutschick, {\it ``Inequivalence of the 
classes of   classical and quantum harmonic potentials: Proof by example''}, 
Phys. Rev. {\bf D  23}  (1981) 922--926.
\bibitem{RWR} R.W. Robinett, {\sl Quantum Mechanics}, Oxford U.P., 1997.
\bibitem{Darboux} G. Darboux, {\it ``Sur  une proposition relative aux \'equations 
lin\'eaires''}, 
Comptes Rendues {\bf 94} (1882) 1456--1459.
\bibitem{AM80} {P.B. Abraham and H.E. Moses},
{\it  ``Changes in potentials due to changes in the point spectrum: Anharmonic oscillators with exact solutions''},
{Phys. Rev.} {\bf A 22}  (1980) 1333--40.
\bibitem{Gelfand-Levitan} B. M. Levitan, {\it ``Sturm-Liouville operators
on the entire real axis with the same discrete spectrum''}, Math. USSR-Sb. {\bf
60} (1988) 77--106. 
\bibitem{mactrub}H.P. McKean and E. Trubowitz, {\it ``The isospectral class of the quantum mechanical harmonic oscillator''},
Commun. Math. Phys. {\bf 82} (1981) 471--495.
\bibitem{PZ98}A.M. Perelomov and Ya. B. Zel'dovich,
{\sl ``Quantum Mechanics: Selected Topics''},
World Scientific (1998).
\bibitem{JoK52} R. Jost and W. Kohn, {\it ``Equivalent potentials"},
Phys. Rev. {\bf 88}  (1952) 382--385.

\end{thebibliography}
\end{document}